\documentclass[12pt,preprint]{aastex}

\shorttitle{MHD shock-cloud interaction}
\shortauthors{Shin et al.}

\begin{document}
\title{The Magnetohydrodynamics of Shock-Cloud Interaction in
Three Dimensions}

\author{Min-Su Shin, James M. Stone, and Gregory F. Snyder}
\affil{Princeton University Observatory, Peyton Hall, Princeton, NJ 08544}

\begin{abstract}
The magnetohydrodynamic evolution of a dense spherical cloud as it
interacts with a strong planar shock is studied, as a model for shock
interactions with density inhomogeneities in the interstellar medium.
The cloud is assumed to be small enough that radiative cooling,
thermal conduction, and self-gravity can be ignored.  A variety of
initial orientations (including parallel, perpendicular, and oblique
to the incident shock normal) and strengths for the magnetic field are
investigated.  During the early stages of the interaction (less than
twice the time taken for the transmitted shock to cross the interior
of the cloud) the structure and dynamics of the shocked cloud is fairly
insensitive to the magnetic field strength and orientation.  However, at
late times strong fields substantially alter the dynamics of the cloud,
suppressing fragmentation and mixing by stabilizing the interface
at the cloud surface.  Even weak magnetic fields can drastically
alter the evolution of the cloud compared to the hydrodynamic case.
Weak fields of different geometries result in different distributions
and amplifications of the magnetic energy density, which may affect
the thermal and non-thermal x-ray emission expected from shocked clouds
associated with, for example, supernovae remnants.
\end{abstract}

\keywords{ISM: clouds --- ISM:kinematics and dynamics --- magnetic fields
--- MHD --- shock waves --- supernova remnants}

\section{Introduction}

The interaction of shock waves with density inhomogeneities (``clouds")
in the interstellar medium (ISM) is thought to be an important dynamical
process in a multiphase medium (e.g., McKee \& Ostriker 1977); for example
it may contribute to mass exchange between the dense and diffuse phases,
and it may trigger gravitational collapse and star formation (Elmegreen
\& Scalo 2004).  This has motivated a number of numerical simulations of the
idealized problem of a planar shock wave interacting with a spherical
cloud to investigate the detailed hydrodynamics in two dimensions (2D)
assuming axisymmetry (Sgro 1975; Woodward 1976; Nittman, Falle, \& Gaskel
1982; Tenorio-Tagle \& Rozyczka 1986; Rozyczka \& Tenorio-Tagle 1987;
Bedogni \& Woodward 1990).  A comprehensive review of the 2D hydrodynamics
of the shock-cloud interaction problem is given by Klein, McKee, \&
Colella (1994, hereafter KMC).

These 2D calculations reveal the cloud is first crushed by the shock, and
then destroyed by a combination of Kelvin-Helmholtz (KH), Richtmyer-Morton
(RM) and Rayleigh-Taylor (RT) instabilities associated with the
vorticity deposited at, and both the impulsive and steady acceleration
of, the density discontinuity at the surface of the cloud respectively.
Since the nonlinear evolution of these instabilities is often different
in three dimensions (3D) compared to 2D, fully 3D simulations of this
model problem are warranted.  Fully 3D hydrodynamical simulations of
a planar shock interacting with a spherical cloud (Stone \& Norman
1992; Klein \& McKee 1994) show that nonaxisymmetric instabilities of
the vortex rings created by spherical clouds serve to further enhance
the destruction rate and mixing between the cloud and postshock gas.
A comprehensive study of the interaction of strong shocks with clouds
with different shapes and orientations (which is only possible in 3D)
was presented by Xu \& Stone (1995, hereafter XS).

Of course the interaction of a planar shock with a uniform density
spherical cloud in hydrodynamics is a considerable simplification of shock
dynamics as it occurs in the ISM.  By considering only planar shocks,
and neglecting post-shock cooling and the self-gravity of the cloud, such
studies can only be used to model the dynamics of small, non-radiative
clouds.  A variety of authors have considered more realistic models,
including (in various combinations) the effect of optically thin radiative
cooling, thermal conduction, self-gravity, and smooth cloud boundaries
(for recent work, see Fragile et al. 2004; Orlando et al. 2005; Nakamura
et al. 2006).  The interaction of planar shocks with multiple spherical
clouds has also been considered (Poludnenko, Frank, \& Blackman 2002)
in order to investigate a more realistic description of the clumpy ISM.

One of the most important extensions to the purely hydrodynamical studies
of the shock-cloud problem discussed above is to include the effects of
magnetic fields using magnetohydrodynamic (MHD) simulations.  Due to
the additional expense and increased complexity of numerical MHD, the
magnetized shock-cloud interaction problem has been less well studied.
The first MHD simulations in 2D axisymmetry were presented by MacLow et
al. (1994, hereafter M94).  In axisymmetry, the only non-trivial initial
field geometry that can be studied is parallel to the shock normal; thus
M94 investigated the effect of changing the strength of parallel fields.
Further 2D simulations, including relativistic electron transport, were
presented by Jun \& Jones (1999).  More recently, Fragile et al. (2005)
has studied the 2D MHD of shock cloud interaction including cooling, and
Van Loo et al. (2007) have studied the formation of strongly magnetized
clouds via shock compression in 2D.  The evolution of a magnetized
dense cloud initially at rest within a supersonic wind was studied by
Jones, Ryu, \& Tregillis (1996).  This wind-cloud interaction problem
produces dynamics that are similar, but not identical, to shock-cloud
interaction.  The wind-cloud interaction problem has been studied in
fully 3D MHD (Gregori et al. 1999; 2000, hereafter G00), albeit at very
limited numerical resolution (no more than 26 zones per cloud radius).
The results of these MHD studies show the magnetic field can have a
dramatic impact on cloud.  In 2D axisymmetry, magnetic fields parallel to
the shock normal suppress RM and KH instabilities, and reduce mixing.
On the other hand, in 3D G00 found that fields perpendicular to the shock
normal were compressed and amplified upstream of the cloud, accelerating
the growth of RT modes and enhancing cloud destruction.

Since the assumption of 2D axisymmetry limits the initial field geometries
that can be considered, it is important to study the MHD shock-cloud
problem in 3D.  In this paper, we present fully 3D MHD simulations of
shock-cloud interaction with a variety of initial orientations of the
magnetic field (parallel, perpendicular, and oblique to the initial shock
normal), and a variety of initial magnetic field strengths ($\beta = 10$,
1, and 0.5, where $\beta$ is the ratio of the gas to magnetic pressure in
the preshock gas).  Advances in computational resources in recent years
allow us to compute the evolution with up to 120 grid points per cloud
radius, well above that used in previous studies, and comparable to the
{\em effective} resolution of adaptive mesh refinement (AMR) simulations
that KMC argued was necessary
for convergence.  We confirm the previous results of M94 for parallel
shocks, namely fields with $\beta \leq 1$ show far less fragmentation and
mixing than occurs in hydrodynamics.  However, we also find the cloud
evolution is very sensitive to the initial field geometry.  Even weak
($\beta=10$) perpendicular and oblique fields produce a substantially
different evolution compared to weak parallel fields, or hydrodynamics.

The organization of this paper is as follows.  We begin by describing
the setup of the problem, our numerical methods, and diagnostics
in \S2.  In \S3 we discuss the convergence of our numerical results.
Most of our results are presented in \S4, while in \S5 we compare and
contrast the evolution with different field geometries and strengths,
and discuss possible observational consequences of our results.  Finally,
we summarize and conclude in \S6.

\section{Method}

To study shock-cloud interaction, we solve the equations of ideal MHD
\begin{eqnarray}
\frac{\partial \rho}{\partial t} +
{\bf\nabla\cdot} [\rho{\bf v}] & = & 0,
\label{eq:cons_mass} \\
\frac{\partial \rho C}{\partial t} +
{\bf\nabla\cdot} [\rho C{\bf v}] & = & 0,
\label{eq:cloud_mass} \\
\frac{\partial \rho {\bf v}}{\partial t} +
{\bf\nabla\cdot} \left[\rho{\bf vv} - {\bf BB} + {\sf P^*}\right] & = & 0,
\label{eq:cons_momentum} \\
\frac{\partial E}{\partial t} +
\nabla\cdot \left[(E + P^*) {\bf v} - {\bf B} ({\bf B \cdot v})\right] & = & 0 ,
\label{eq:cons_energy} \\
\frac{\partial {\bf B}}{\partial t} +
{\bf\nabla} \times \left({\bf v} \times {\bf B}\right) & = & 0,
\label{eq:induction}
\end{eqnarray}
where $\rho$, ${\bf v}$, and ${\bf B}$ are
the total mass density, velocity, and magnetic field respectively,
and ${\sf P^*} = P + B^{2}/2$ and $E$ is the total energy density
\begin{equation}
  E = \frac{P}{\gamma -1} + \frac{1}{2}\rho v^{2} + B^{2}/2.
\label{eq:total_energy}
\end{equation}
with $\gamma=5/3$.  These equations are written in units such that
the magnetic permeability $\mu=1$.  The color variable $C$ is used to
track mixing between the ambient and cloud material.  We initialize the
problem with $C=1$ within the cloud (see the next section), and $C=0$
everywhere else.  During the resulting evolution, regions in which
$0 \leq C \leq 1$ contain a mixture of cloud and ambient material.
Equation \ref{eq:cloud_mass} can be thought of as a conservation law
for the cloud mass $M_{cl} = \int_{V} \rho C dV$.

\subsection{Initial Conditions}

Although most previous studies of shock-cloud interaction used 
a density discontinuity at the cloud surface (including the 3D hydrodynamical
calculations presented in XS), for 
small clouds it is more realistic to represent the density
distribution with a smooth profile at the boundaries.  Following
previous authors (Kornreich \& Scalo 2000; Nakamura et al. 2006), the density 
profile is given by
\begin{equation}
\rho(r) =  \rho_{0} + \frac{(\rho_c - \rho_0)}{1 + (r / r_{c})^n},
\label{eq:cloud}
\end{equation}
where $\rho_{0}=1$ is the density of the ambient medium,
$\rho_{c}=10$ is the density at the center of the cloud, and the
core radius $r_{c}=0.62$.  Thus, the
cloud is overdense compared to its surroundings by a factor $\chi =
\rho_{c}/\rho_{0} = 10$.  Larger values of $n$ give steeper
density profiles at the surface; we adopt $n=8$ for all calculations
presented in this paper.

The density distribution defined in equation \ref{eq:cloud} extends to
infinity.  Thus, we must adopt an arbitrary definition for the boundary
of the cloud.  In this paper, we define the boundary at $r=r_{b}$,
where $\rho (r_{b}) = 1.002\rho_{0}$.  For the parameters given above,
this gives $r_{b} = 1.77$.  Thus, the total column density of the cloud
in these units is
\begin{equation}
  \Sigma_{cl} =  \int_{-r_{b}}^{r_{b}} \rho(r) \approx 15
\end{equation}  
The value of $\Sigma_{cl}$ is important to the dynamics because we expect
significant evolution to occur once the cloud has swept up a comparable
column of the postshock gas.

Initially the cloud is in pressure equilibrium with its surroundings.
We set the gas pressure $P=1$ everywhere, giving a sound speed in the
ambient gas of $C_{s} = \sqrt{5/3}$.

We study three different initial directions and strengths for the initial
magnetic field.  Depending on the orientation of the magnetic field
in the preshock gas with respect to the unit normal perpendicular to
the shock front, our simulations use either parallel, perpendicular, or
oblique magnetic fields.  The latter use fields inclined at $45^{\circ}$
to the shock normal.  The strength of the magnetic field can be defined
by a plasma $\beta$ calculated in the preshock medium.  For each field
geometry, we study fields with a strength given by $\beta=10$, 1 and 0.5

The final quantity required to specify the initial conditions is the sonic
Mach number of the incident shock, $M_{s}$.  We study the evolution of strong
shocks with $M_{s}=10$ in this paper.  Given $M_{s}$, and the magnetic
field orientation and strength, it is straightforward to calculate
the properties of the postshock gas from the shock jump conditions.
The evolution of stronger shocks should be the same as presented here
with the appropriate dimensionless scaling (see M94).  The evolution of
much weaker shocks may be different, but is not considered further here.

\subsection{Numerical Methods and Grid}

All of the numerical simulations presented in this paper use Athena,
a newly developed Godunov scheme for astrophysical MHD.  Details
of the algorithms implemented in Athena are given in Gardiner \& Stone
(2005; 2007).  These algorithms have been extended to integrate 
equation \ref{eq:cloud_mass} along with the usual equations of ideal MHD.

The computational domain covers the region $ -3.66 \leq x \leq 6.34$, $-5 \leq
y \leq 5$, and $-2.5 \leq z \leq 2.5$ for all of the simulations.  The cloud is initially centered
on the origin.  The shock is located at $x=-2.66$, and propagates along
the $x-$axis.  The initial orientation of the magnetic field 
in the preshock gas is along
the $x-$ axis for parallel shocks, along the $y-$axis for perpendicular
shocks, and at $45^{\circ}$ in the $x-y$ plane for oblique shocks.

Once the shock impacts the cloud, the cloud is rapidly
accelerated, and would leave the computational domain through the right-hand
boundary in the $x-$direction in only a few dynamical times.  To keep the
shocked cloud centered in the grid, we compute the mass-averaged cloud
velocity
\begin{equation}
 \langle v_{x} \rangle = \frac{1}{M_{cl}} \int_{V} \rho C v_{x} dV
\label{eq:x-velocity-moment}
\end{equation}
every time-steps, and integrate the MHD equations in a frame of reference
moving at $\langle v_{x} \rangle$.
This keeps the center of the cloud at rest with respect to
the grid.  Essentially, our domain moves with the cloud, considerably
reducing the computational volume needed to follow the evolution.  We report
all our results with respect to the initial frame of reference which is at
rest with respect to the preshock medium.

All of the simulations presented in \S4 use a grid of $N_x \times N_y
\times N_z = 680 \times 680 \times 340$, or about 160 million zones.
This is thirty times larger than the 3D hydrodynamic shock-cloud
interaction simulation first presented in Stone \& Norman (1992).
This resolution results in roughly 120 grid points covering the cloud
radius $r_b$.  To investigate the convergence of our results (\S3), we
have also computed the evolution on a grid with one half the resolution,
or $N_x \times N_y \times N_z = 340 \times 340 \times 170$ zones.

\subsection{Time Scales}

The time taken for the incident shock to propagate across the {\em
interior} of the cloud defines the ``cloud-crushing time" $t_{cc}$, a
characteristic time scale for the shock-cloud interaction problem (KMC).
For a cloud with the density profile equation \ref{eq:cloud}, the cloud-crushing
time is defined as
\begin{equation}
t_{cc} \equiv \frac{r_{c}}{v_s} = \frac{\chi^{1/2} r_{c}}{M_s C_s},
\end{equation}
where $v_{s}$ is the shock velocity in the cloud interior.
We will report all of our results with time given in units of $t_{cc}$.
Generally, our simulations are run until $t \sim 20 t_{cc}$.

For the cloud parameters given in \S2.1, $t_{cc} \sim 0.15$. To convert to
typical values of small clouds in the ISM, if we assume $r_{c} = 4000~{\rm
AU}$, $C_s = 10^5~{\rm cm ~ s^{-1}}$ for $\chi = 10$ and $M_s = 10$, then
$t_{cc}\sim 6,000$ years.  The physical size of the computational domain
in this case is $\sim 0.3~\rm{pc} \times 0.3~\rm{pc} \times 0.15~\rm{pc}$.

\subsection{Diagnostics}

There are a variety of diagnostic quantities we will use to analyze our
results.  We define the cloud-mass-weighted average of any quantity $f$ as
\begin{equation}
\langle f \rangle = \frac{1}{M_{cl}} \int \rho C ~ f ~ dV , 
\end{equation}
For example, the average cloud velocity defined in equation
\ref{eq:x-velocity-moment} is just the cloud-mass-weighted average
of the $x-$component of velocity.

To follow changes in the shape of the cloud, the
mass-weighted moment along the $x-$axis is used (MacLow et al. 1994)
\begin{equation}
a = [5 ( \langle x^2 \rangle - \langle x \rangle^2 )]^{1/2}, 
\end{equation}
The expressions for the moments along the $y-$ and $z-$axis, $b$ and $c$
respectively, are similar.  For the parallel shock simulations, we expect
symmetry about the $x-$axis, so that $b = c$.  This can be used as a code
check.

The acceleration of the cloud can be studied using the mass-weighted
velocity along each axis, e.g. $\langle v_{x} \rangle$ (already given
by equation \ref{eq:x-velocity-moment}), $\langle v_{y} \rangle$ and
$\langle v_{z} \rangle$.

In order to trace mixing between the cloud and ambient material, we use
the mixing fraction $f_{mix} \equiv M_{mix}/M_{cl}$, where following XS we
define $M_{mix}$ to be the total mass of cells in which the color variable
$0.1 \leq C \leq 0.9$.  Such cells are those in which the cloud and ambient
gas are well mixed.

Finally, the RMS of velocity along each orthogonal axis is useful
to understand mixing and the generation of turbulence (Nakamura et al. 2006).
The root-mean-square (RMS) of the $x-$component of the velocity is
\begin{equation}
\delta v_x = (\langle v_{x}^{2}\rangle - \langle v_{x}\rangle^2)^{1/2}
\end{equation}
with similar expressions for the RMS along the $y-$ and $z-$axes,
$\delta v_y$ and $ \delta v_z$ respectively.
We also expect symmetry in $\delta v_y$ and $ \delta v_z$ for the parallel
shock case.

\section{Convergence Study}

As with any numerical study, it is important to confirm that diagnostic
quantities measured from the simulations are converged, that is they
do not change with increased resolution.  Not all aspects of the flow
will be converged.  For example, the nonlinear growth and saturation
of MHD instabilities at the cloud interface generates
turbulence.  Without explicit dissipation to fix the Reynolds and
magnetic Reynolds numbers in the flow, increasing the resolution will
lead to ever smaller scales in this turbulence.  Thus, we should expect
quantities that are sensitive to turbulence at small scales (such as
the mixing rate between cloud and ambient gas) may not be converged,
while quantities which are insensitive to turbulence at small scales
(such as the shape of the cloud) are more likely to show convergence.
We begin be investigating which of the diagnostics in our calculations
show convergence.

Since the shock-cloud interaction problem has been so well studied in
the past, the criteria for convergence in hydrodynamic simulations
has been investigated by many
other authors.  For example,  KMC used AMR methods to achieve the highest
effective resolution possible in 2D simulations at that time (about $10^{2}$ grid points
per cloud radius), and argued that such resolution were necessary to see
convergence of the solutions.  More recently, Nakamura et al. (2006) have
repeated this study by measuring the decrease in errors in
various diagnostic quantities (such as the size of the cloud axes $a$, $b$,
and $c$, and the $\delta v_{x}$ and $\delta v_{y}$
defined in \S2.4) with increasing resolution.  Again, they conclude that
around $10^2$ grid points per cloud radius is necessary for convergence in
the hydrodynamical problem in 2D.

In figure 1, we plot the time evolution of the axis ratio $a/b$, the
RMS of velocity in the $x-$ and $y-$directions $\delta v_{x}$ and
$\delta v_{y}$, and the mixing
fraction $f_{mix}$, for a parallel shock simulation with $\beta=0.5$,
at resolutions of both $680 \times 680 \times 340$ and $340 \times 340
\times 170$ zones.  As in previous studies (KMC, Nakamura et al. 2006),
we find the first three of
these quantities are converged, with virtually no change in the curves
between the two resolutions.  Moreover, the relative errors between these
quantities are similar to those reported in Nakamura et al. (2006) for
2D hydrodynamic simulations at the same resolution.  However, the last of
these quantities $f_{mix}$ clearly does not show convergence, with about
a 20\% decrease in this quantity at late times at the highest compared
to the lower resolution.  From the discussion above, this result is not
surprising.  Without explicit diffusion, the fraction of mixed cells, and
therefore $f_{mix}$, continues to decrease with increasing resolution.
Indeed, for immiscible fluids, the fraction of mixed fluid should be zero
at infinite resolution, consistent with the trend visible in figure 1.

Thus we conclude that most aspects of the MHD shock-cloud interaction
problem are well converged at the resolutions presented here, at least
as well as most previous 2D hydrodynamic studies of this problem.
The important exception is the mixing fraction.  Instead, the mixing
fraction observed in our calculation corresponds to that associated
with a small but non-zero mass diffusion at the grid scale mediated by
numerical effects.  Thus, although this implies the mixing fraction
we report is probably not meaningful in an absolute sense, the {\em
relative} change in the mixing rate between different simulations with
the identical resolution, but different magnetic field strengths or
geometries can still be a useful diagnostic, since they all have the
same effective numerical diffusion at the grid scale.  In this case,
{\em differences} between $f_{mix}$ are likely due to the differences
in the dynamics rather than numerical effects at the grid scale.

\section{Results}

Table 1 lists the parameters of the simulations discussed in this paper.
A total of nine simulations have been run, all of which have used the
same grid and resolution as described in \S2.2.  Each run is labeled by
the initial field geometry (PA for parallel shocks, PE for perpendicular,
and OB for oblique) and field strength ($\beta$).  We discuss the results
from each of these geometries in turn in the following subsections.

\subsection{Parallel shocks}

Figure 2 presents volumetric renderings of the cloud density $\rho C$ at
four times during the evolution of the weak and strong parallel fields,
runs PA10 and PA05 respectively.  The image is computed viewing along
the direction given by the unit normal $\hat{\bf n} = (0.5,0.866,0)$,
that is at $30^{\circ}$ to the $y-$axis, and is centered on the origin.
We expect the shock to have propagated across the diameter of the cloud by
$t=2t_{cc}$.  Thus, at $t=3.76t_{cc}$ shown in the first panel in figure
2, both the incident and transmitted shocks are well past the cloud,
the initial compression phase of the cloud has ended, and the shocked
cloud material is now re-expanding.  Note there is little difference
in the structure of the cloud between the weak and strong field cases
at this time, since the mechanical energy of the shock rather than the
magnetic energy of the cloud dominates.  The only significant difference
at this time is that the column of compressed material behind the head of
the cloud is broader for $\beta = 0.5$, since the strong field resists
compression towards the axis by the incident shock as it is refracted
around the cloud.  However, at later times the structure of the cloud in
the weak and strong field cases begins to show significant differences.
For the weak field, the tip of the cloud is shredded into a complex
network of filaments due to shear instabilities associated with the
vorticity deposited at the cloud surface, similar to the hydrodynamic
problem in 3D (SN, XS).  On the other hand, for a strong field the cloud
resists fragmentation.  Shear instabilities cause wrinkles, fingers,
and clumping at the cloud surface, however to a much smaller degree
in comparison to the weak field.  Direct comparison of figure 2 with,
e.g. figure 2 in XS, shows that even for the weak field case, the
destruction of the cloud is decreased in comparison to hydrodynamics.

Previous studies (Hawley \& Zabusky 1989) have shown that the vortensity
$\omega^{2} \equiv (\nabla \times {\bf v})^{2}$ is a useful diagnostic
for the hydrodynamic shock-cloud interaction problem.  Figure 3 plots
volumetric renderings of $\omega^{2}$ for the weak and strong parallel
fields at the same four times and using the same viewing angle, as
shown in figure 2.  Again, the evolution with a weak field is similar,
but not identical, to the hydrodynamic case (SN, XS).  At the earliest
time, a thin layer of vorticity has been deposited at the cloud surface,
while the reflection of the incident shock on-axis behind the cloud has
generated a very strong vortex ring downstream of the cloud.  This ring is
completely absent in the strong field case.  The thin sheet of vorticity
is associated with this ring marks the location of the incident shock,
which has been curved by passage over the cloud.  (By comparing figures
2 and 3, note the disk of cloud material behind the head of the cloud
is downstream of the location of the incident shock, marked by this thin
sheet of vorticity.)  At later times, KH instability in the
sheet of vorticity at the head of the cloud results in the formation of
a complex network of tangled vortex filaments that produce the complex
density structures at the corresponding times in figure 2.  In contrast,
for the strong field case the vortex sheet remains largely intact, indicating
the KH instability is suppressed.

The stabilizing effect of a strong magnetic field for parallel
shocks is evident in the time evolution of each component of the
velocity RMS.  Figure 4 shows the evolution of the parallel and
transverse components, $\delta v_{x}$ and $\delta v_{\perp} \equiv (\delta
v_{y}^{2} + \delta v_{z}^{2})^{1/2}$ respectively, normalized by the {\em
preshock} sound speed for runs PA10, PA1, and PA05.  The largest values
of $\delta v_{x}$ occurs very early, at about one $t_{cc}$.  At this
time, the transmitted shock is only half way through the cloud, giving
a large dispersion between the shocked and unshocked cloud material.
The largest values of the transverse component of the dispersion occurs
slightly later, at slightly less than $2t_{cc}$.  This corresponds to
the period of maximum transverse compression by the transmitted shock.
Thereafter, the longitudinal component decreases uniformally, with only
a small difference between different values of $\beta$ (with the largest
$\beta$ having the largest dispersion).  The transverse component also
decreases to a constant value, however this value depends strongly on
the field strength.  For $\beta = 10$, $\delta v_{\perp}/C_{s} \approx
0.5$, whereas for $\beta = 0.5$, $\delta v_{\perp}/C_{s} \approx 0.2$.
The longitudinal component of the velocity RMS is indicate of
uniform acceleration of the cloud, whereas the transverse component is
more indicative of postshock turbulence.  This turbulence is subsonic
with respect to the preshock gas (and therefore highly subsonic with
respect to the postshock gas).  The large decrease in the transverse
velocity RMS with decreasing $\beta$
is indicative that strong fields suppress turbulence.

We expect the evolution of the velocity RMS and magnetic field
to be tightly coupled.  Figure 5 plots the time evolution of the energy
associated with amplification of both the longitudinal field 
$(B_{x}^2 - B_{x,0}^{2})/B_{x,0}^{2}$, and the transverse
field $(B_{y}^{2} + B_{z}^{2})/B_{x,0}^{2}$ in a simulation box (where $B_{x,0}^{2}$ 
is the energy associated with the initial field), for each value of
$\beta$.  The weak field $\beta=10$ shows the largest amplifications in
both plots.  Moreover the longitudinal field is amplified significantly
more than the transverse.  This reflects the importance of compression
of the longitudinal field on-axis to form a flux rope behind the cloud
(M94) in comparison to turbulent amplification of the transverse field.
The field geometries produced by the shock-cloud interaction will be
discussed further in \S5.

Finally, the relative changes in the time evolution of the mixed
fraction $f_{mix}$ for different values of $\beta$ is shown in figure 6.
The largest mixing occurs for $\beta=10$, while both of the stronger
field cases ($\beta = 1$ and 0.5) are comparable.  Since $f_{mix}$
is not converged (as discussed in \S3), we assign no relevance to the
particular values of this parameter in any run.  However, the decrease in
the {\em relative} values with increasing field strength is meaningful,
and reflects the decrease in fragmentation and mixing of the cloud
evident in figure 2.

\subsection{Perpendicular shocks}

Figure 7 shows 3D renderings of the cloud in runs PE10, PE1, and PE05 at
three times during the evolution.  We have found that it is difficult
to interpret the structure of the cloud in perpendicular shock runs
with strong fields from volumetric renderings as shown in figure 2
(similarly, we find plotting vortensity is not useful in this case).
Instead, figure 7 shows a combination of the $\rho C=6$ isosurface, a
slice of $\rho C$ plotted on the horizontal plane $y=0$, and magnetic
field vectors plotted on the vertical plane $z=0$.  The viewing angle
and focus point in each panel is identical to that in figure 2.

At the earliest time, $t=3.76t_{cc}$, the structure of the cloud is
similar in all three runs (and remarkably similar to the parallel shock
runs shown in figure 2).  There is a slight asymmetry in the cloud
shape along the directions perpendicular and parallel to the field,
but this is clearly evident only in the $\beta=0.5$ case.  At later
times, however, the cloud structure with different field strengths is
significantly different.

For the weak ($\beta=10$) field, the cloud is fragmented, with a clear
asymmetry in the density structures between the $x-y$ and $x-z$ planes.
Since initially the magnetic field lies in the $x-y$ plane, vortical
motion in that plane twists the field, and thus the field resists such
motion.  However, in the $x-z$ plane, vortical motion simply interchanges
field lines, and so the field does not affect such motion.  Thus, strong
vortices form in the $x-z$ plane, with cloud material entrained at the
center of the vortices.  The magnetic field vectors shown in the $x-y$
plane demonstrate that the field lines drape over the surface of the
cloud as expected, and that the field is strongly amplified (as shown
by the color and length of the arrows) as it is stretched downstream of
the cloud.

For both of the stronger ($\beta=1$ and 0.5) fields, the cloud no longer
becomes fragmented at late times, but instead resembles a sheet aligned
with the field.  In these two cases, stretching of the field over the
cloud is evident at early times ($t=3.76t_{cc}$ and $t=7.97t_{cc}$).
However, at late times the field is vertical everywhere.  This occurs
because acceleration of the cloud to the postshock velocity is rapid
(e.g. for $\beta=0.5$, the postshock density and $v_{x}$ is about 3.54
and 9.58 respectively; this means the cloud sweeps up its own column
density by $t \approx 0.5 t_{cc}$, and by a few $t_{cc}$ the cloud has
accelerated to the postshock flow speed).  Once the cloud is at rest with
respect to the postshock gas, the longitudinal perturbations in the field
propagate away from the cloud as shear Alfv\'{e}n waves, restoring the
field to purely vertical near the cloud.  These waves are clearly evident
at $t=7.97t_{cc}$ for the $\beta=1$ case.  Thus, unless the column density
of the cloud is significantly higher (so that the cloud takes much longer
to accelerate), there will not be strong longitudinal fields near the
cloud surface at late times.  Instead, strong longitudinal fields will
be associated with the shear Alfv\'{e}n waves far from the cloud surface.

Previous 3D studies of wind-cloud interaction using perpendicular fields
(Gregori et al. 2000) discussed the importance of field compression
at the leading surface of the cloud, forming a ``magnetic bumper"
which accelerates fragmentation.  At $t=3.76t_{cc}$ there are small
amplitude vertical striations in the leading surface of the cloud in
both the $\beta=1$ and 0.5 runs, associated with MHD RT instabilities
in the compressed field upstream of the cloud. The
growth rate of the MHD RT instability is given by (Chandrasekhar 1961)
\begin{equation}
\sqrt{a k \frac{\rho_{2} - \rho_{1}}{\rho_{2} + \rho_{1}} - \frac{B^{2} k^{2}}{2 \pi (\rho_{1} + \rho_{2})}}, 
\end{equation}
where $a$ is the effective acceleration of the cloud,
$\rho_{1}$ and $\rho_{2}$ are the densities of the ambient gas and cloud
respectively, $B$ is the magnetic field strength, and 
$k$ is the wave number of the perturbation.  By measuring from the
simulations the effective
acceleration of the cloud, the densities across the surface of the cloud, the
magnetic field strength, and the wavelength of the perturbations shown in
figure 7, we find the growth time from the 
above equation is about 10 times shorter than the acceleration time.  Thus,
we expect the MHD RT instability to grow rapidly.  By $t=7.97t_{cc}$ these
striations have grown to large amplitudes, fragmenting the cloud into
vertical columns.  Even the $\beta=10$ begins to show the striations at
this time.  However, as discussed in \S5, at late times the strongest
amplified field is still associated with filaments downstream of the
cloud, rather than the compressed field upstream.

The anisotropic shape of the cloud produced by a perpendicular field is
best illustrated by the time evolution of the ratio of the moments of the
cloud axes, $b/a$ and $c/a$ respectively.  Figure 8 plots these moments
for all three field strengths, runs PE10, PE1, and PE05 respectively.
Initially both ratios show a peak between $1-2t_{cc}$ associated with
the decrease in $a$ as the cloud is crushed along the direction of
propagation of the shock.  Thereafter the ratios decrease to constant
values.  Note for strong fields $b/a > c/a$, indicating the cloud is
extended along the direction of the background field.  For the weak field
($\beta=10$) the ratios are nearly equal and very similar to the values
for the parallel shock shown in figure 1.

The evolution of the magnetic energy associated with the $x-$ and
$z-$components of the magnetic field for runs PE10, PE1, and PE05 is shown
in figure 9.  The evolution of $B_{y}^2$ is dominated by linear growth
due to compression as the shock propagates across the grid, and therefore
is not shown.  Once again the weakest field shows the largest growth.
Moreover, the energy associated with the longitudinal field $B_{x}^2$
shows much larger growth than that associated with the transverse field
$B_{z}^2$.  This latter asymmetry simply reflects the fact the field is
initially orientated in the $x-y$ plane, thus $B_{x}$ can be created by
bending the field over the cloud surface (vortical motion in the $x-y$
plane), while $B_{z}$ cannot.  The large decrease in both $B_{y}^2$ and
$B_{z}^2$ at late times for $\beta=1$ and 0.5 is caused by the propagation
of the shear Alfv\'{e}n waves through the transverse boundaries of the grid.
For example, it is evident in figure 9 that the largest values of $B_{x}$
are associated with these waves, and once they leave the grid the field
becomes almost vertical everywhere, so $B_{x}$ drops to nearly zero.
Note that no decrease is seen for the weak field $\beta=10$ run.  In this
case, field amplification is caused by winding up in vortices, and in
this case the tension in the amplified field is never large enough to
cause unwinding.

Finally, figure 10 plots $f_{mix}$ for runs PE10, PE1, and PE05.
Once again, a strong decrease in the relative amounts of mixing is
observed as the field strength increases, a factor of two between the
$\beta=10$ and $\beta=0.5$ cases.  While the {\em absolute} values of
$f_{mix}$ are not converged, the {\em relative} decrease is real, and reflects
the decrease in turbulence and fragmentation with stronger fields.

\subsection{Oblique shocks}

Figure 11 plots the $\rho C=6$ isosurface, a slice of $\rho C$ plotted
on the horizontal plane $y=0$, and magnetic field vectors plotted on
the vertical plane $z=0$ for the oblique field simulations, runs OB10,
OB1, and OB05 respectively.  The viewing angle and focus point for the
$\beta=10$ run OB10 is identical to that used in previous figures,
while for the $\beta=1$ and 0.5 plots the plots are centered on the
point $(x,y,z) = (0,0,-1)$.

Once again, we find that at early times ($t=3.76t_{cc}$), there is little
difference in the structure of the cloud with different field strengths
(or by comparison to figures 2 and 7, for different field geometries).
However, at later times, the cloud in the weak field run becomes highly
fragmented, whereas with stronger oblique fields the cloud is more
sheet-like.  As with perpendicular shocks, small amplitude striations
are evident in the leading surface of the cloud with strong fields; they
indicate the presence of the MHD RT instability associated with compressed
upstream field.  At later times, the instability grows to large amplitude,
and fragments the cloud into vertical columns.  However, the most striking
aspect of the evolution is that the cloud becomes significantly displaced
from the $y=0$ plane with strong fields.  For oblique shocks, $v_{y}$
is no longer zero in the postshock gas, instead it varies from -0.022
($\beta=10$) to -0.434 ($\beta=0.5$).  Thus, the cloud is simply dragged
in the negative $y-$direction by the postshock flow, while retaining a
sheet-like structure aligned with the field.

Fragmentation of the cloud in the weak field ($\beta=10$) oblique shock
case is more complex than for perpendicular shocks.  In addition to the
suppression of vortical motion in the $x-y$ plane, oblique fields also
affect vortical motion in the $x-z$ plane.  Thus, while the $\beta=10$
perpendicular shock tended to fragment the cloud into ribbons and sheets
in the $x-z$ plane, the oblique shock tends to fragment the cloud into
isolated structures.

For the strong field runs, magnetic tension tends to erase longitudinal
perturbations, and near the cloud the field is aligned with the direction
in the postshock gas (about $15^{\circ}$ from the $y-$axis for both
$\beta=1$ and $\beta=0.5$).  Strong longitudinal perturbations in the
field propagate away from the cloud as shear Alfv\'{e}n waves.  Figure 12
plots the time evolution of the magnetic energy associated with the $x-$
and $z-$components of the perturbed magnetic field for all three oblique
field runs.  Again, we do not show $B_{y}^2$ since it is dominated by
shock compression.  The evolution of the magnetic energies is similar
to the perpendicular shock case as shown in figure 9.  The strongest
amplification is for the weakest fields, and the energy associated with
the longitudinal field $B_{x}^{2}$ is much larger than that associated
with the transverse field $B_{z}^{2}$.

Other diagnostics of the cloud evolution for oblique shocks are very
similar to that shown for perpendicular shocks, and so are not shown here.
For example, the evolution of the shape of the cloud as measured by the
axis ratios $b/a$ and $c/a$ is similar to that shown in figure 8; the
cloud is elongated in the $y-$direction and so $b/a$ tends to be largest.
In addition, the mixing fraction $f_{mix}$ evolves in a similar manner
to the perpendicular field case (shown in figure 10).  In particular,
the relative amount of mixing decreases strongly with increased field
strength, indicating the suppression of turbulent fragmentation and
mixing with strong oblique fields.

\section{Discussion}

By studying the MHD shock-cloud problem in 3D, we have for the first time
been able to compute the evolution of spherical clouds with transverse
magnetic fields.  At early times ($t \leq 4t_{cc}$), the evolution is
dominated by the mechanical energy of the shock, and the structure of
the cloud is independent of the magnetic field strength or geometry.
However, at late times, and for strong fields, the morphology of the
cloud is substantially different depending on the initial field geometry
(e.g. compare the structure of the density for runs PA05, PE05, and OB05
shown in figures 2, 7, and 11 respectively).  For strong parallel fields,
the cloud becomes a flattened disk aligned perpendicular to the field.
However, for strong perpendicular and oblique fields, the cloud becomes
more sheet-like, aligned with the field.

Although the fact that at late times the cloud morphology depends on
the field geometry for strong fields is not surprising, we find that
even for weak fields, the structure of the density and magnetic field
is substantially different depending on the initial field geometry.
Figure 13 shows volumetric renderings of $\rho C$ at $t=7.97t_{cc}$
for the parallel, perpendicular, and oblique shocks with weak fields
(runs PA10, PE10, and OB10 respectively).  Also shown are volumetric
renderings of the magnetic energy $B^{2}$ at the same time for these
three runs.  For the parallel shock case, the density at the tip of
the cloud is strongly affected by MHD instabilities, and already shows
a filamentary appearance.  The magnetic energy is dominated by a strong
``flux-rope" behind the cloud, formed by lateral compression of the mean
field by refraction of the incident shock.  This structure was discussed
extensively in the 2D MHD simulations of M94.  The field at the tip of the
cloud is weak and filamentary.  In contrast, for both the perpendicular
and oblique shocks, the density distribution at the head of the cloud is
much smoother, and dominated by a ring just downstream of the head of
the cloud.  Apparently even a weak transverse field can suppress the RM
(Wheatley, Pullin, \& Samtaney 2005), KH (Ryu, Jones, \& Frank 2000)
and RT (Stone \& Gardiner 2007a; 2007b) instabilities at the tip of the
cloud that are present in the parallel shock case.  The magnetic energy
in the perpendicular and oblique shocks is also substantially different
than the parallel shock.  Now, most of the energy is associated with a
smooth cap draped over the head of the cloud, and the single flux-rope behind
the cloud becomes two parallel filaments.

The changes in the morphology of the density and magnetic energy with
field geometry are important because of the  potential observational
consequences.  Both thermal x-ray emission from shock-heated cloud
material, and non-thermal synchrotron emission from shock-accelerated
particles, is expected from shock-cloud interaction.  Recently, several
authors have modeled thermal x-ray emission from hydrodynamic simulations
of shocked clouds relevant to supernovae remnants (Miceli et al. 2006;
Orlando et al. 2006), finding that thermal conduction and evaporation
of cloud material can have important effects.  Since the geometry of
the magnetic field can strongly affect thermal conduction, we expect
significant differences between the thermal emission from the different
weak field simulations shown in figure 13.  For example, thermal
conduction should be much more important in the parallel shock case.
The sheath of amplified magnetic field that drapes over the cloud in the
perpendicular and oblique shocks may significantly reduce conduction.
Modeling the non-thermal synchrotron emission requires MHD models to
follow field amplification.  Synthetic maps of the emission for the
parallel shock case were computed by M94, most of the emission was
associated with the strong flux rope generated downstream of the cloud.
It is clear from figure 13 that the synchrotron emission from the
perpendicular and oblique field shocks will be quite different, with
strong emission from the sheath at the head of the cloud, as well as
the twin flux tubes downstream.

Another observational diagnostic of shock-cloud interaction
is provided by studies in the optical.  Through detailed comparison of
Balmer-dominated emission filaments in an isolated shocked cloud in the
Cygnus Loop with 2D hydrodynamical simulations, Patnaude \& Fesen (2005)
have found the density profile at the cloud surface is most likely smooth,
so that turbulent striping is suppressed.  However, the cloud density
structures shown in figure 13 also show that weak transverse fields suppress
turbulent stripping.  It is possible that MHD simulations with weak transverse
magnetic fields also will be a good fit to the optical data.

\section{Conclusion}

We have presented well-resolved 3D simulations of the interaction of
$M_{s}=10$ planar shocks with magnetized, spherical clouds.  By performing
the simulations in 3D, we are able to capture the nonaxisymmetric modes of
instabilities that contribute to the destruction of the cloud, and more
importantly, we are able to study general field geometries, including
initially parallel, perpendicular, and oblique to the shock normal.
Our primary conclusions are the following:
\begin{enumerate}
\item In the early stages of the interaction ($\leq 4 t_{cc}$), magnetic fields make almost no difference to the structure of the shocked cloud.

\item For strong parallel fields, the shocked cloud 
has a disk-like structure at late times, flattened along the direction of
propagation of the shock.  For strong perpendicular and oblique fields,
the shocked cloud is sheet-like at late times, orientated parallel to the
postshock magnetic field.

\item For weak fields, the late stages of evolution are
dominated by turbulent stripping and fragmentation, regardless of the
initial field geometry.  However, even with weak fields, the morphology of the
cloud and distribution of magnetic energy is substantially different depending on the
initial geometry (see figure 13).

\item Weak subsonic turbulence is generated by the interaction.  Strong
fields substantially decrease the amplitude of the turbulence.

\end{enumerate}

The focus of this paper has been the detailed MHD of shock-cloud
interaction.  For this reason, we have adopted a number of simplifying
assumptions, which should be relaxed in future work.  Perhaps the most
important is the neglect of radiative cooling and thermal conduction.
Fragile et al. (2005) have shown that cooling can affect the MHD of
shock-cloud interactions, leading to the formation of very dense knots
that can survive longer.  Moreover, detailed modeling of the x-ray
emission from shocked clouds requires a proper treatment of both cooling
and conduction.  Computational resources and numerical algorithms are
now capable of modeling MHD shock cloud interaction in fully 3D with
the relevant physics, future studies aimed at directly modeling observed
shock interactions in the ISM would be very fruitful.

\acknowledgements
We thank Tom Gardiner to his many contributions to the Athena code
used in this work.  Computations were performed on the IBM Blue Gene
at Princeton University, and on computational facilities supported by
NSF grant AST-0216105.  JS thanks the DOE for financial support through
DE-FG52-06NA26217, and NASA for support through NNG06GJ17G.



\begin{table}[t]
\caption{Parameters of simulations.}
\begin{tabular}{ccc} \hline \hline \\
Model & Initial Field Orientation & $\beta$ \\ \hline \\
PA05 & (2, 0, 0) & 0.5 \\
PA1 & ($\sqrt{2}$, 0, 0) & 1 \\
PA10 & ($\sqrt{0.2}$, 0, 0) & 10 \\ \\ \hline \\
PE05 & (0, 2, 0) & 0.5 \\
PE1 & (0, $\sqrt{2}$, 0) & 1 \\
PE10 & (0, $\sqrt{0.2}$, 0) & 10 \\ \\ \hline \\
OB05 & ($\sqrt{2}$, $\sqrt{2}$, 0) & 0.5 \\
OB1 & (1, 1, 0) & 1 \\
OB10 & ($\sqrt{0.1}$, $\sqrt{0.1}$, 0) & 10 \\
\end{tabular}
\end{table}


\begin{figure}[t]
\plotone{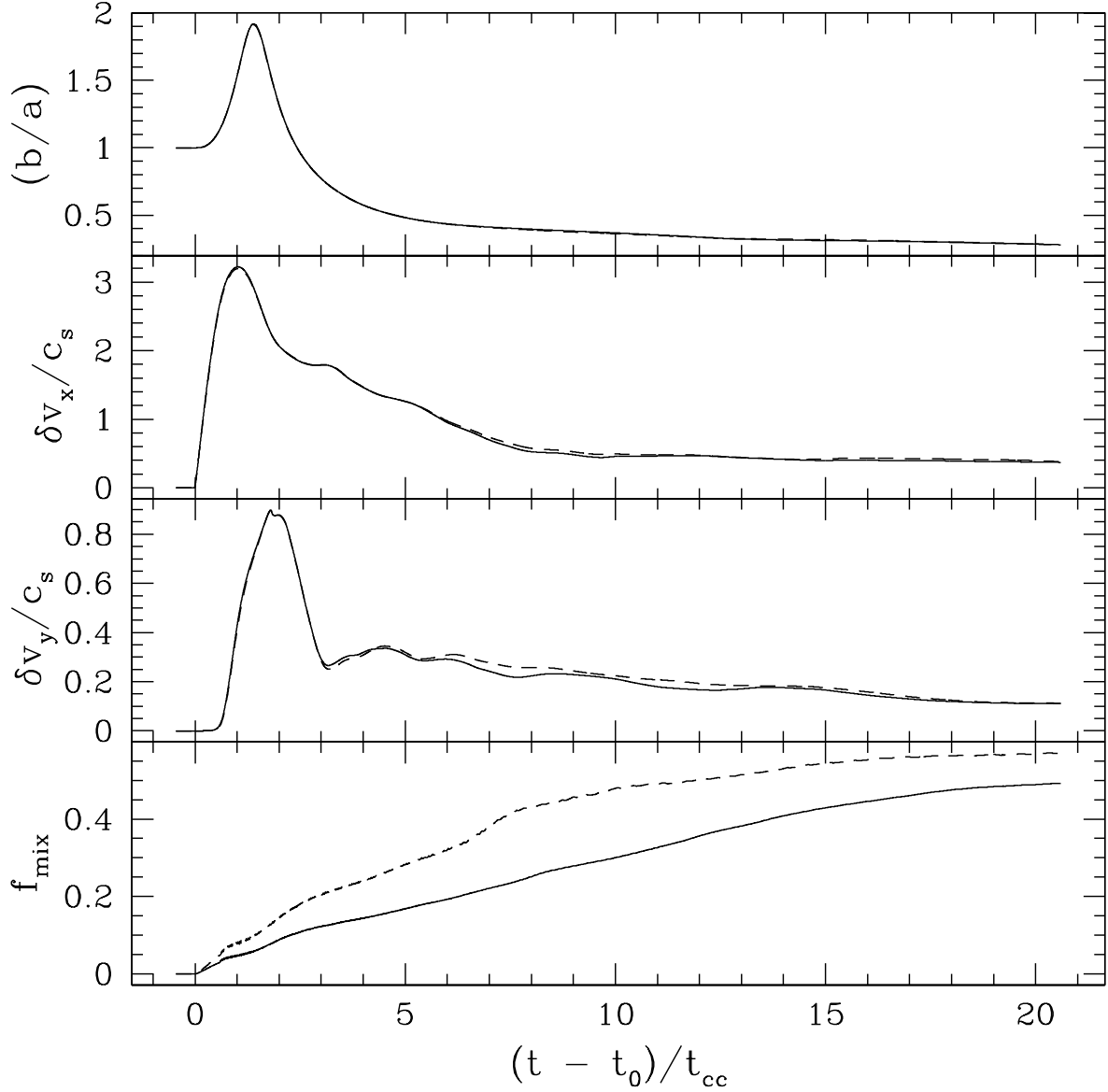}
\figcaption{Evolution of the axis ratio $b/a$, velocity RMS 
in the $x-$ and $y-$directions $\delta v_{x}$ and $\delta v_{y}$,
and mixing fraction $f_{mix}$ for a parallel shock
with $\beta=0.5$ at numerical resolutions of $680\times680\times340$
(solid line) and $340\times340\times170$ (dashed line).
All quantities except the mixing fraction show convergence.
}
\end{figure}

\begin{figure}[t]
\epsscale{0.9}
\plotone{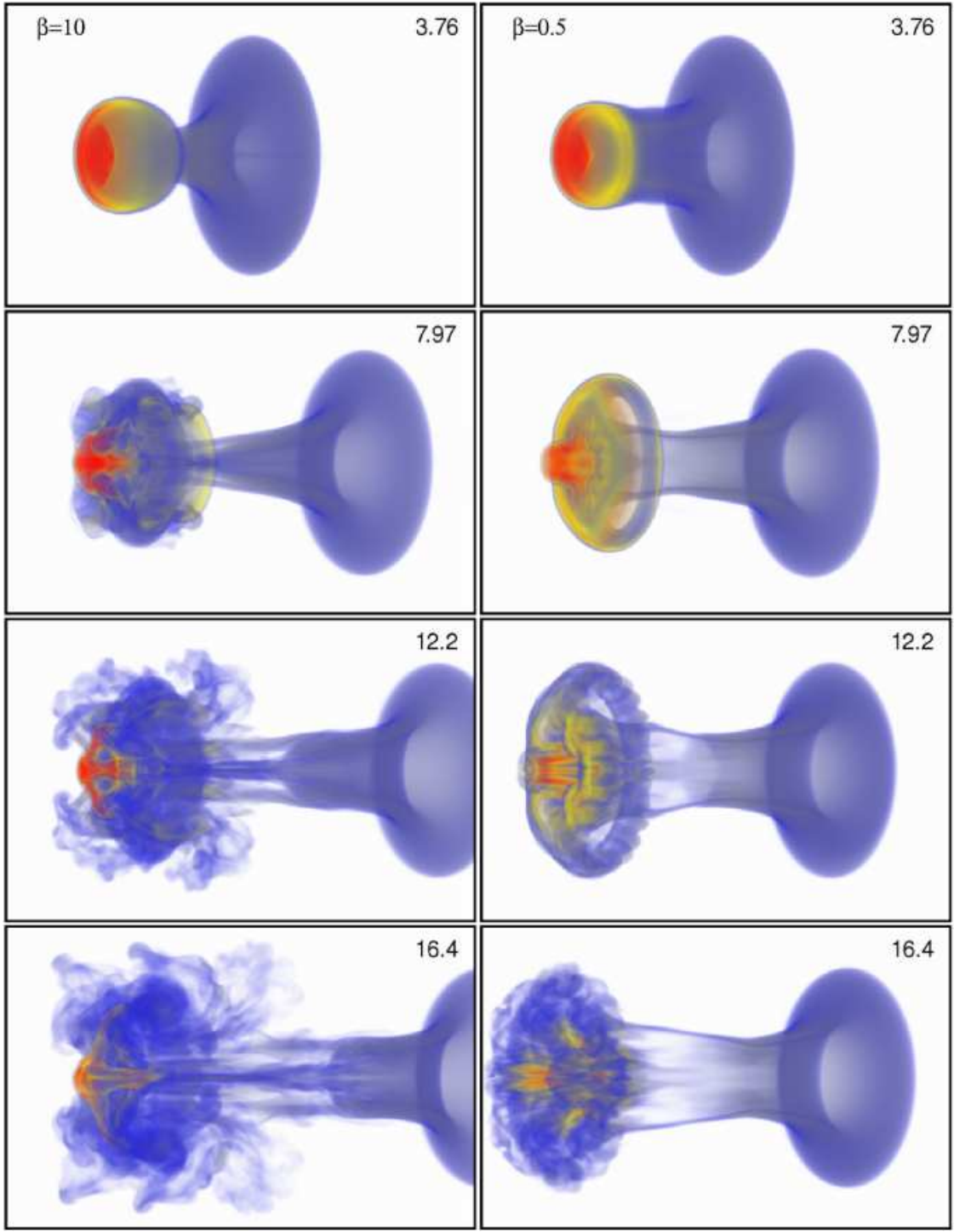}
\figcaption{Volumetric renderings of the cloud density for both the weak
field ($\beta=10$, left column) and strong field ($\beta=0.5$, right column)
parallel shock simulations.  The time of each image is shown in the upper
right corner of each panel, in units of the cloud-crushing time $t_{cc}$,
and measured from the moment at which the shock first impacts the cloud.
}
\end{figure}

\begin{figure}[t]
\epsscale{0.9}
\plotone{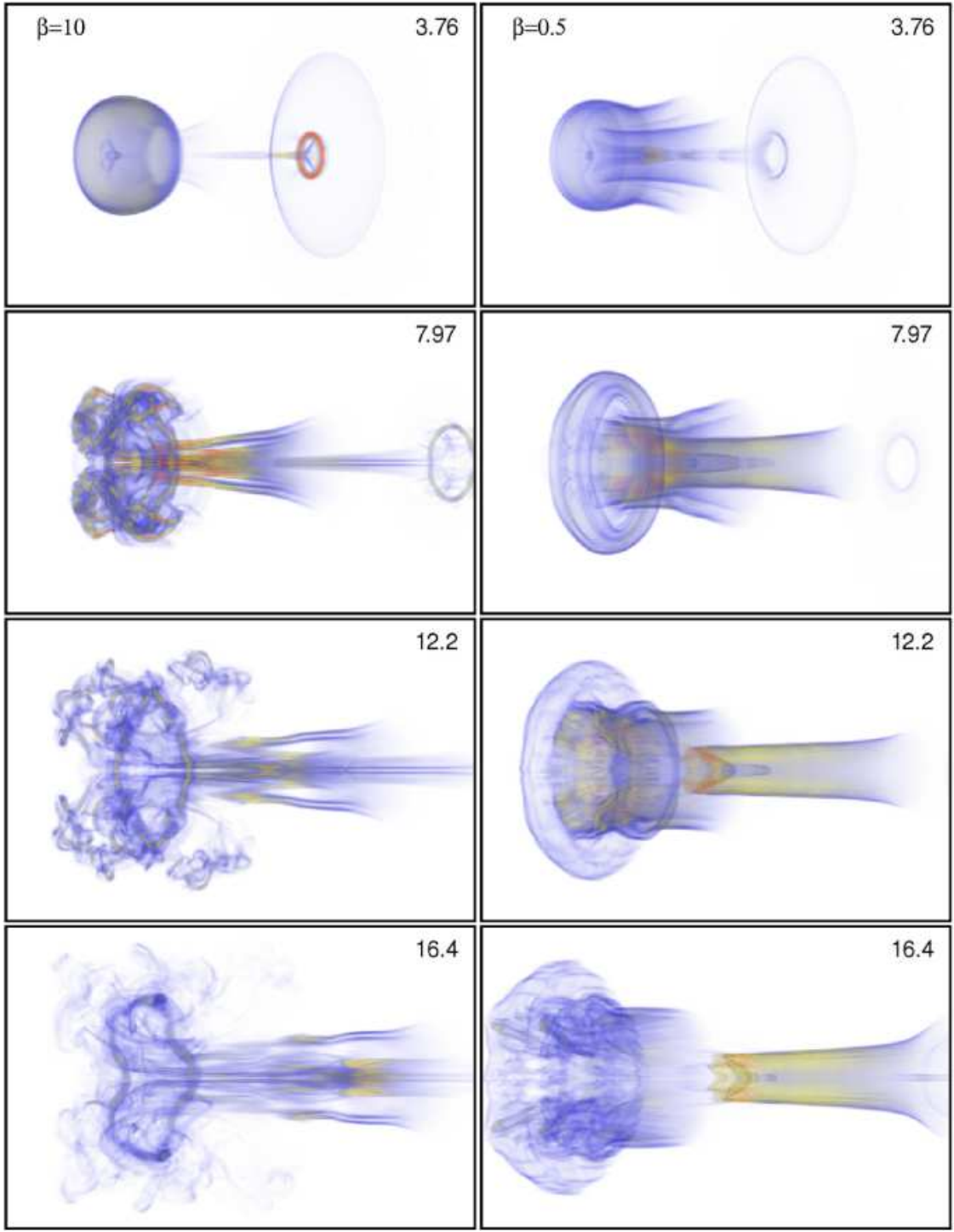}
\figcaption{Volumetric renderings of the vortensity $\omega^{2} =
(\nabla \times {\bf v})^2$ for both the weak field ($\beta=10$,
left column) and strong field ($\beta=0.5$, right column) parallel shock
simulations.  The time of each image is shown in the upper right corner,
as in figure 2.
}
\end{figure}

\begin{figure}[t]
\epsscale{0.8}
\plotone{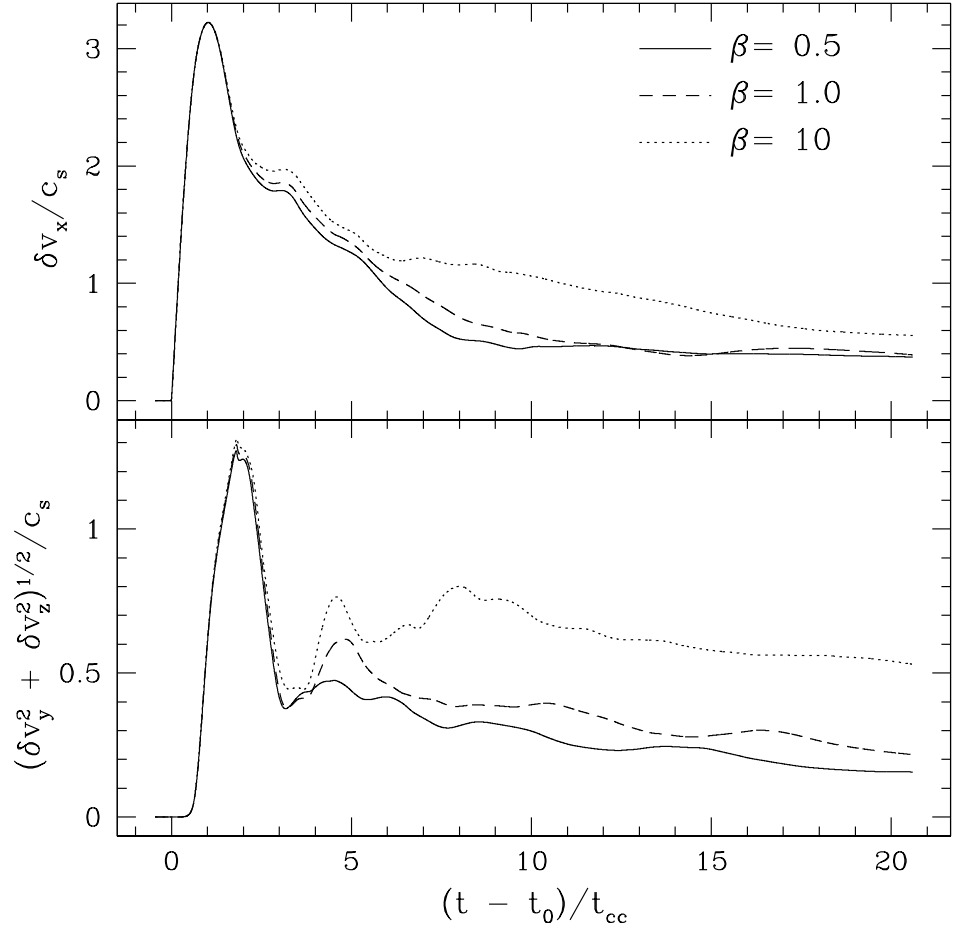}
\figcaption{Evolution of the velocity RMS parallel (top panel) and
perpendicular (bottom panel) to the shock normal for different initial magnetic
field strengths, for the parallel shock simulations.
}
\end{figure}

\begin{figure}[t]
\epsscale{0.4}
\plotone{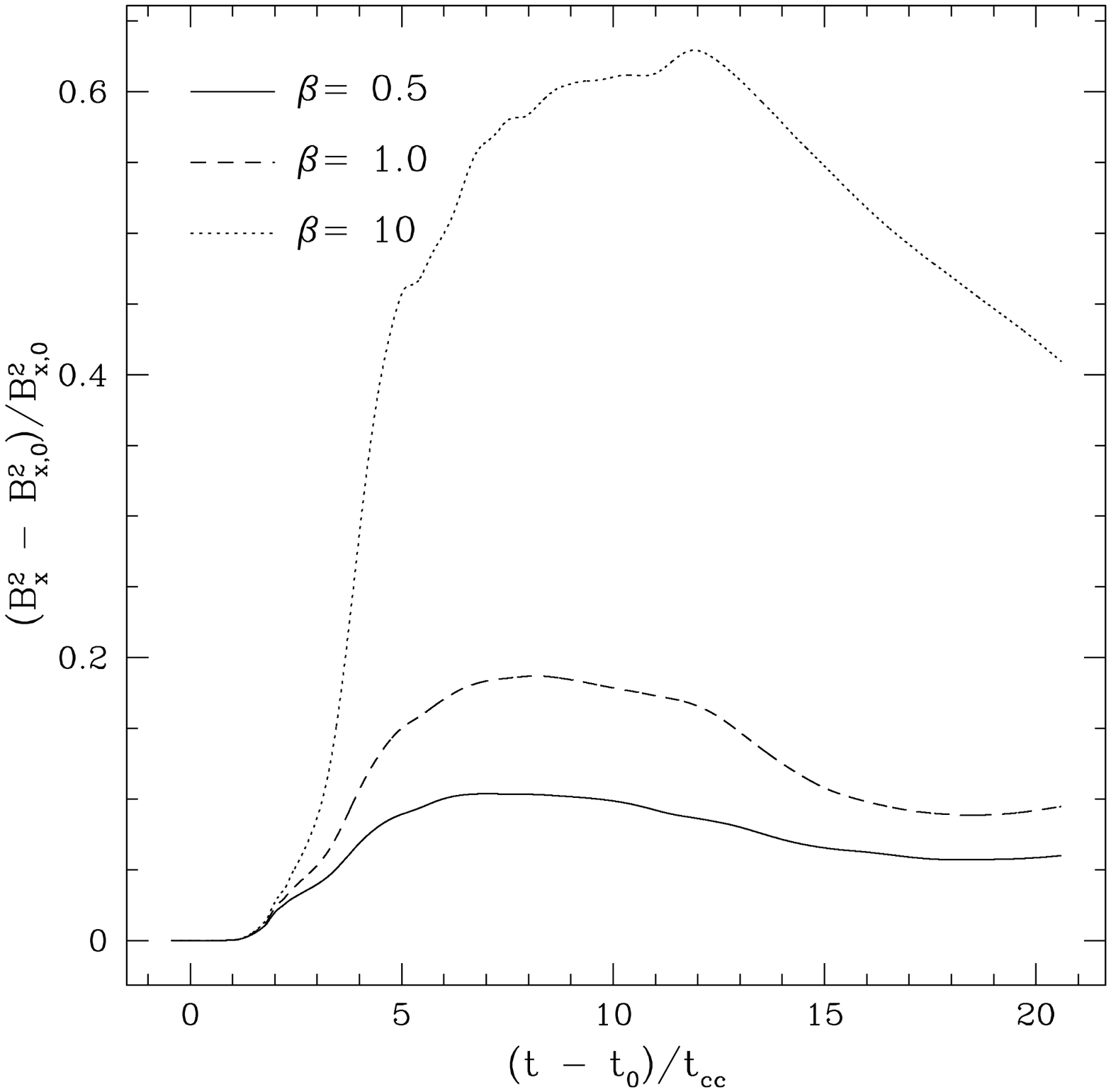}
\plotone{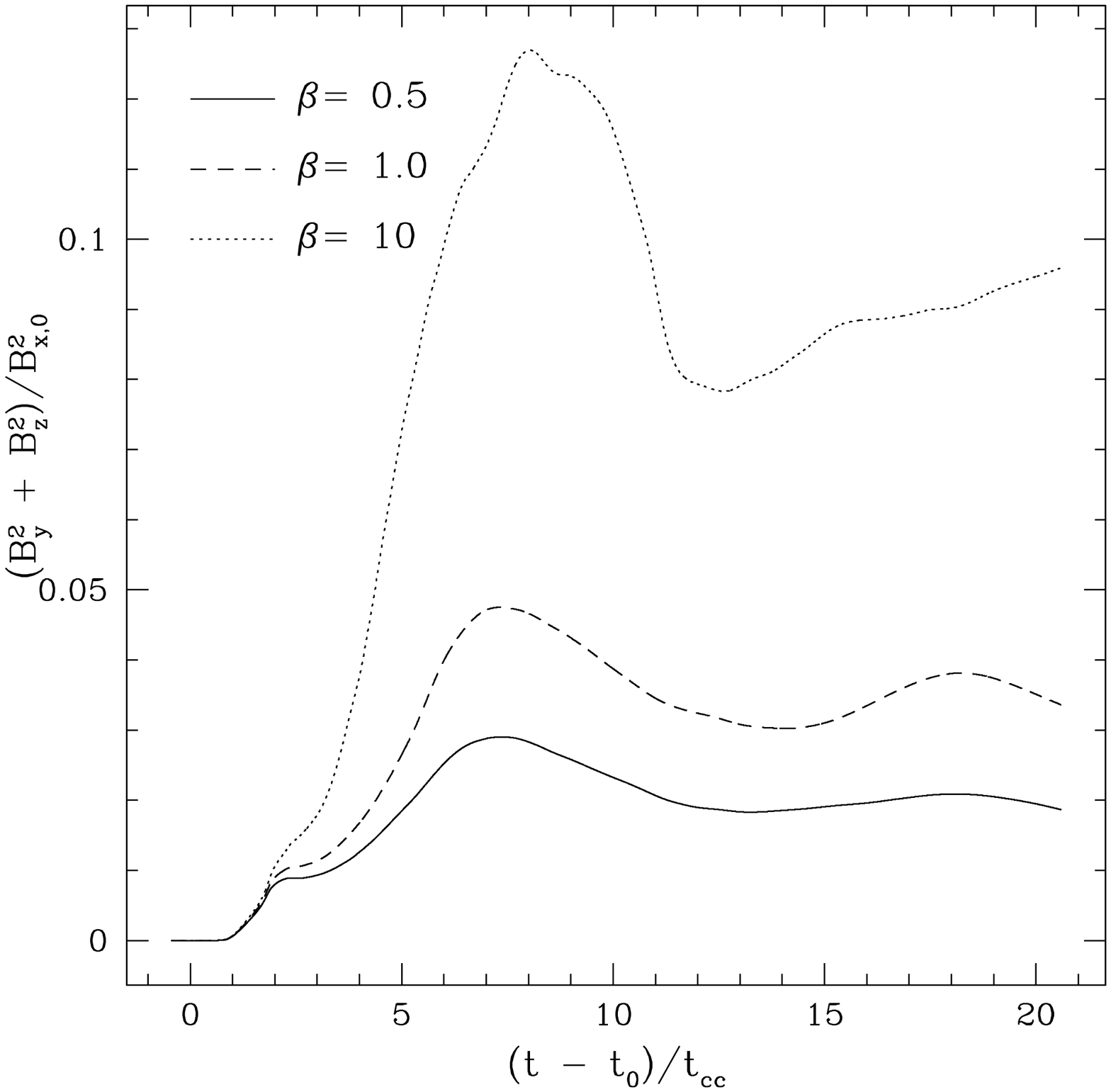}
\figcaption{Evolution of the change in magnetic energy density for different
$\beta$ for the parallel shock simulations, measured in units of the energy
density in the initial field $B_{x,0}^{2}$.  The left panel shows the change 
associated with the parallel component of the field, and the right with
the perpendicular component. We note that the decrease after $\sim 12 
{\rm (t - t_{0})/t_{cc}}$ is the result of a small amount of cloud material 
that flows out of the simulation box.
}
\end{figure}

\begin{figure}[t]
\plotone{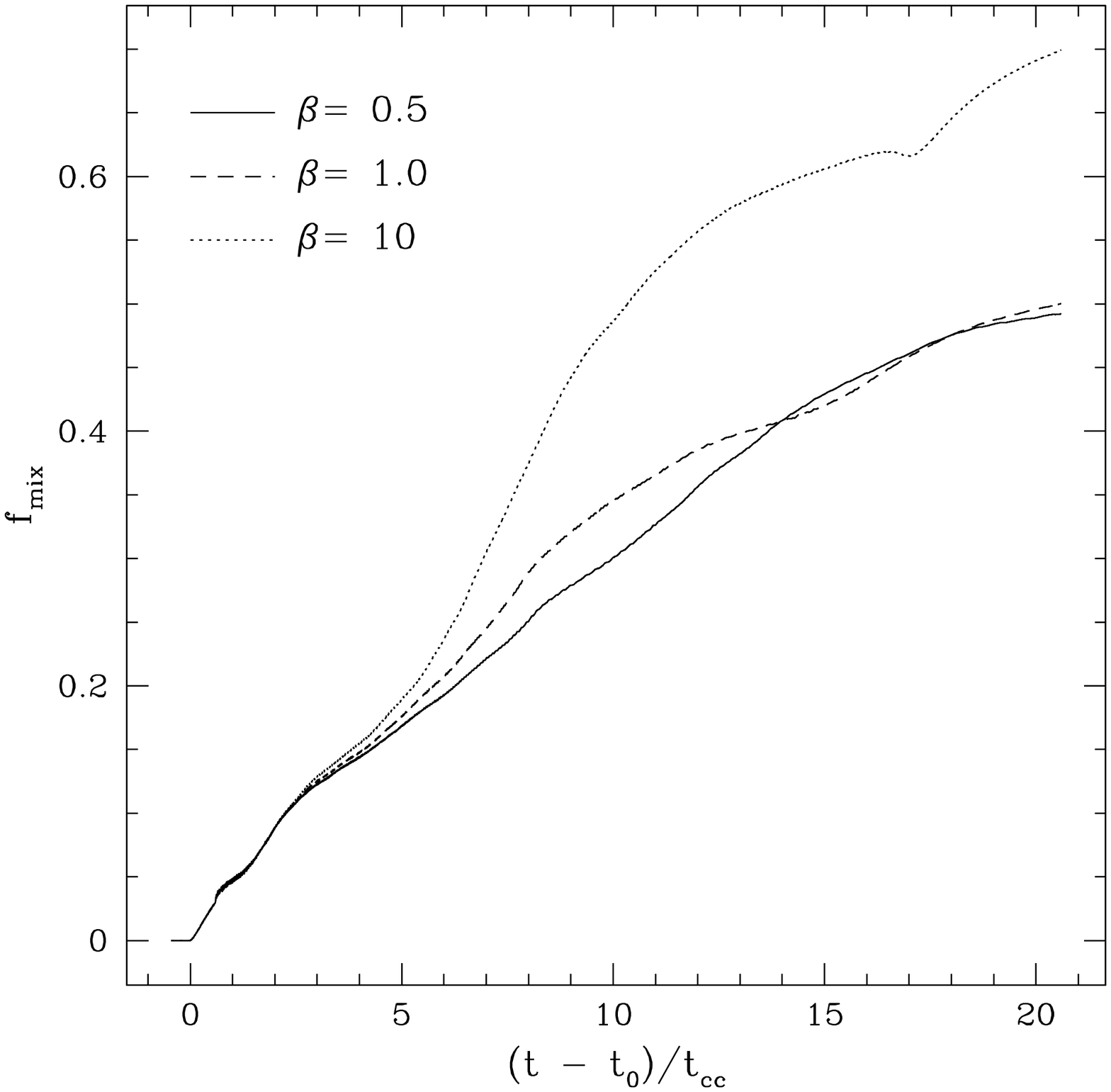}
\figcaption{Evolution of the mixing fraction for parallel shocks with
different initial magnetic field strength, measured by the initial $\beta$.
}
\end{figure}

\begin{figure}[t]
\epsscale{1.0}
\plotone{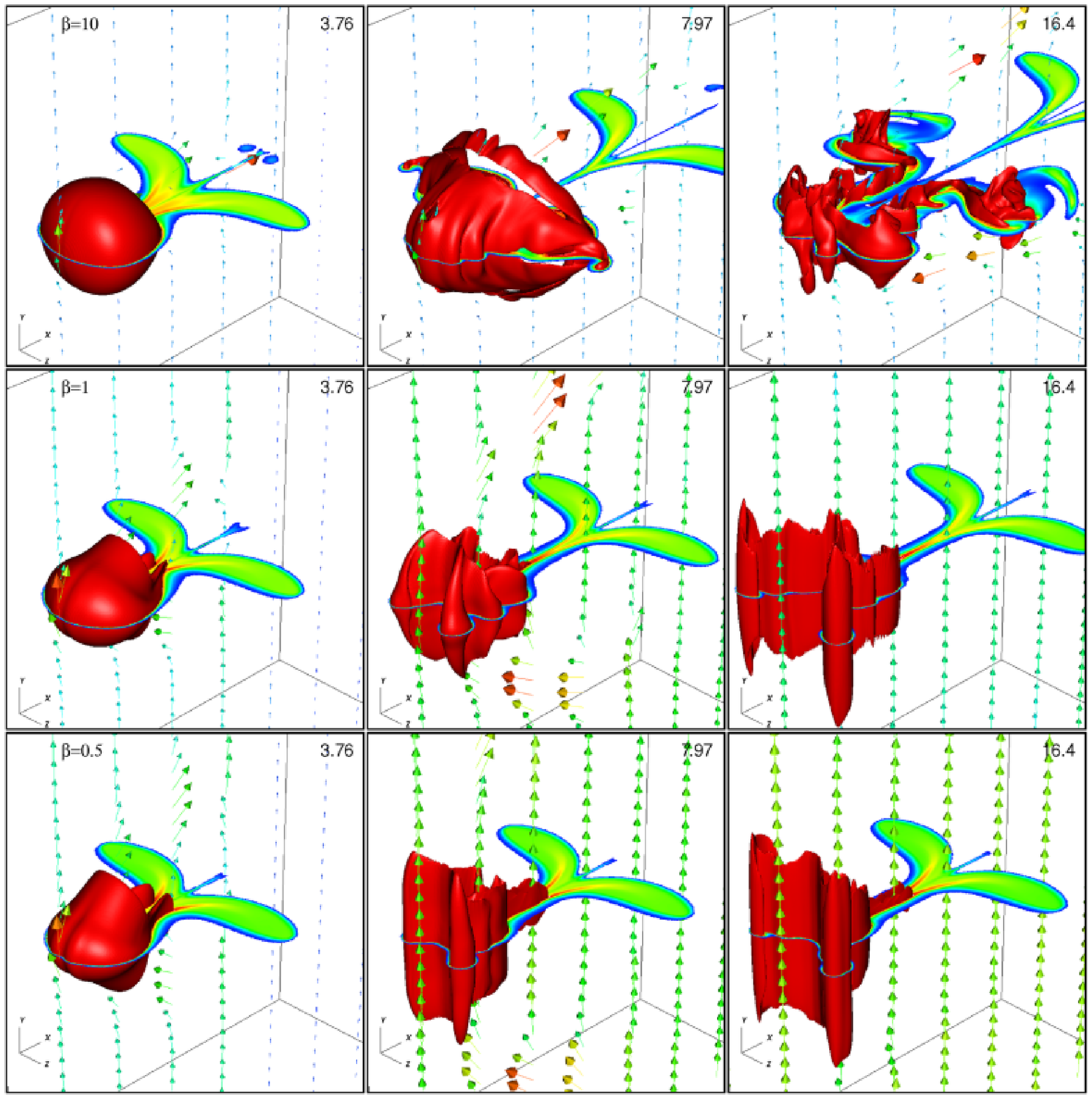}
\figcaption{Isosurface and horizontal slice at $y=0$ of the cloud mass density
$\rho C$, and magnetic field vectors (scaled and colored by 
$\parallel {\bf B} \parallel$)
on a vertical slice at $z=0$, for the
perpendicular shock simulations with an initial field that is
either weak ($\beta=10$, top row), equipartition ($\beta=1$, middle row),
or strong ($\beta=0.5$, bottom row).
}
\end{figure}

\begin{figure}[t]
\epsscale{0.8}
\plotone{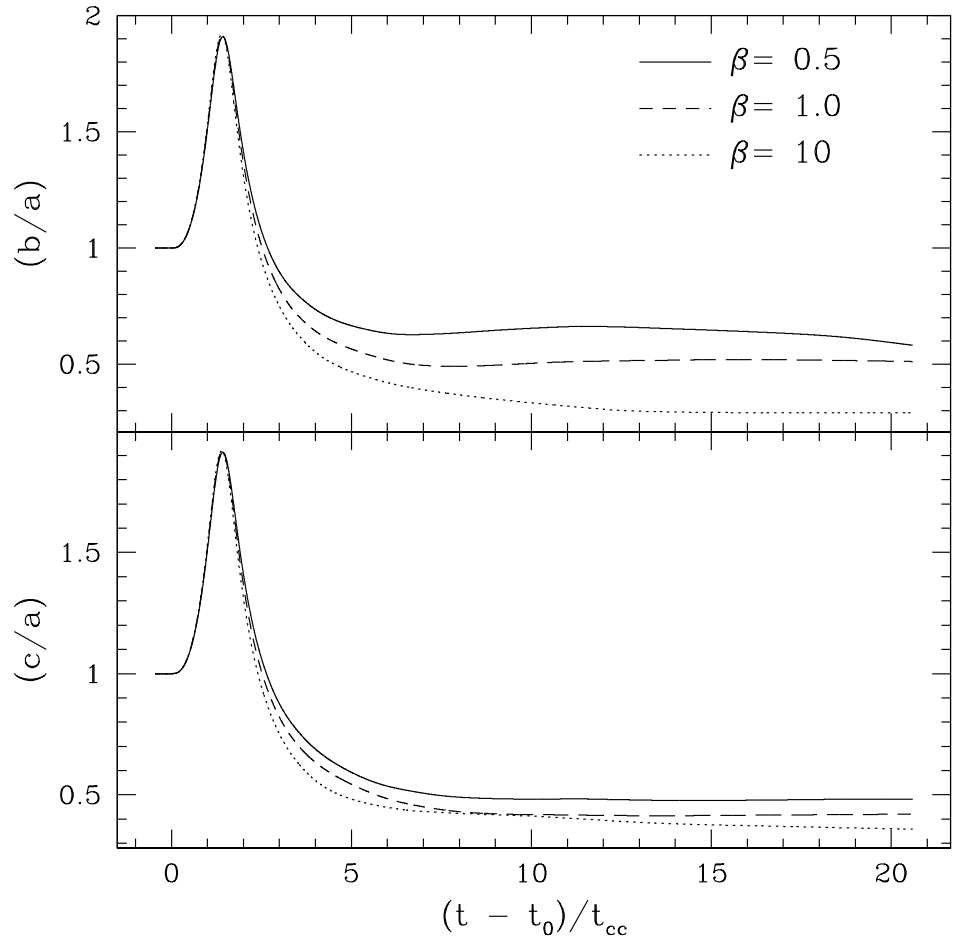}
\figcaption{Axial ratios $b/a$ (top panel) and $c/a$ (bottom panel)
with different magnetic field strengths for perpendicular shocks.
}
\end{figure}

\begin{figure}[t]
\epsscale{0.4}
\plotone{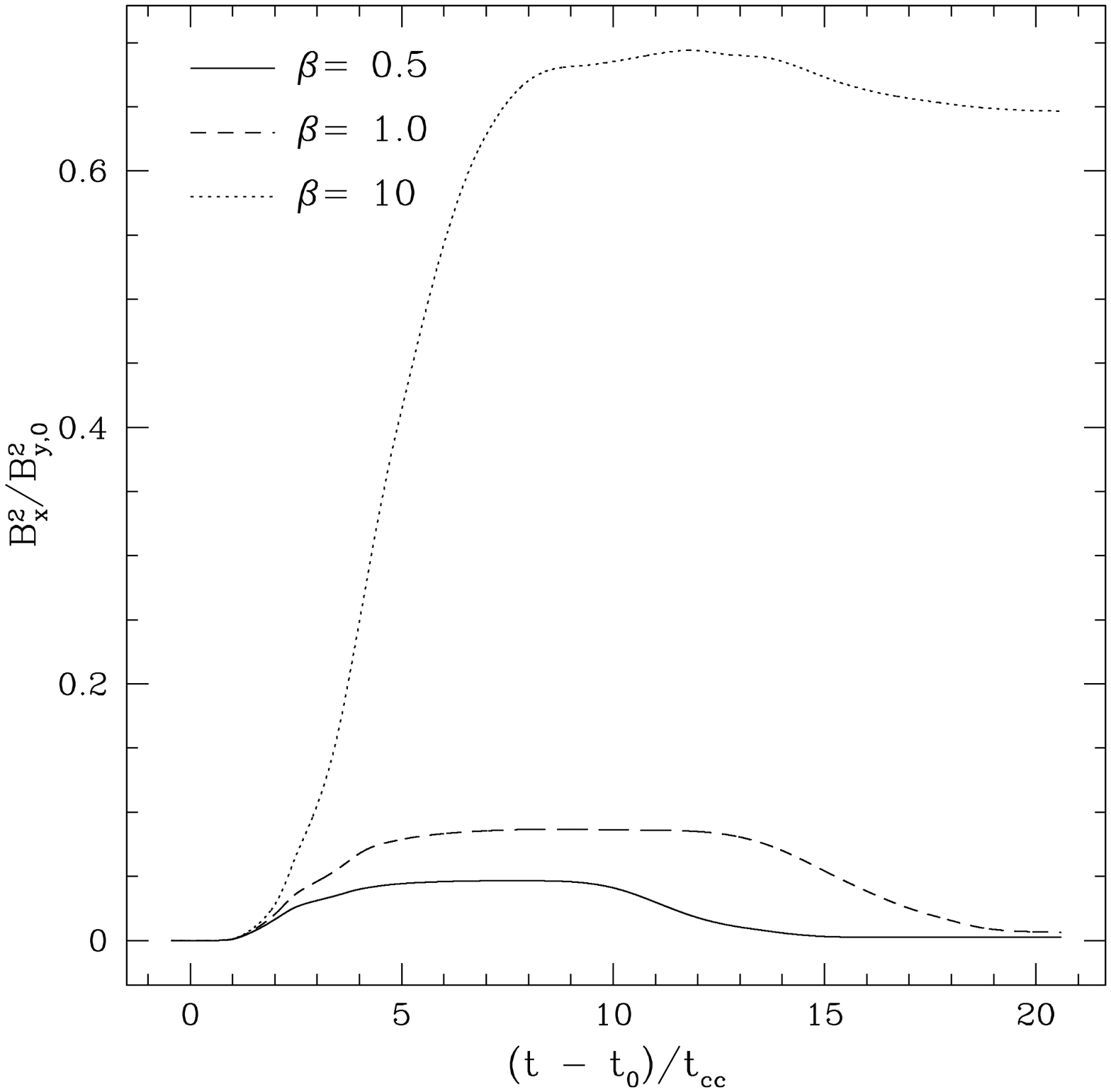}
\plotone{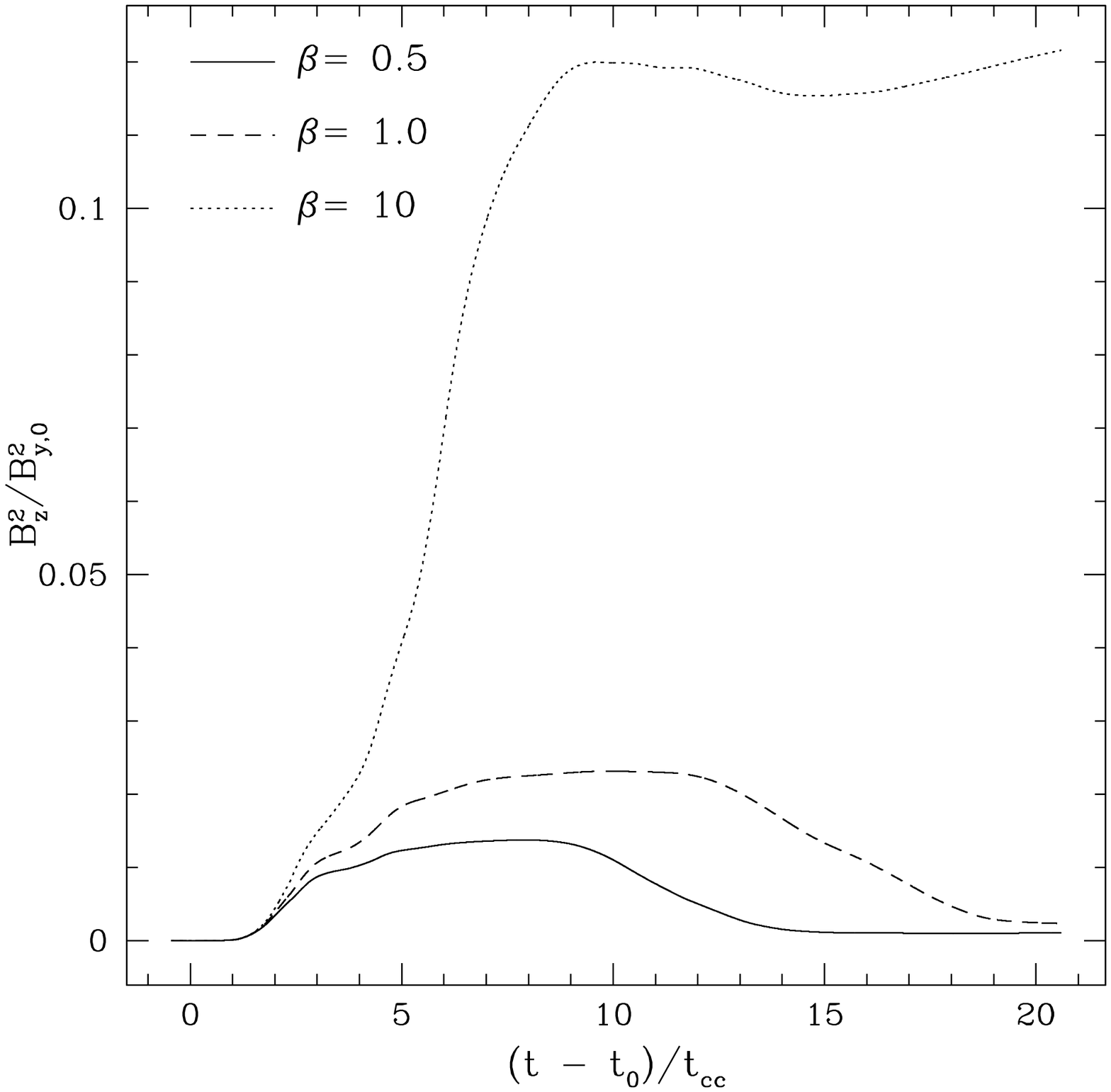}
\figcaption{Evolution of the change in magnetic energy density for different
$\beta$ for the perpendicular shock simulations, measured in units of the energy
density in the initial field $B_{y,0}^{2}$.
The evolution of $B_y^2$ is dominated by amplification due to
shock compression, and so is not shown. 
}
\end{figure}

\begin{figure}[t]
\epsscale{0.8}
\plotone{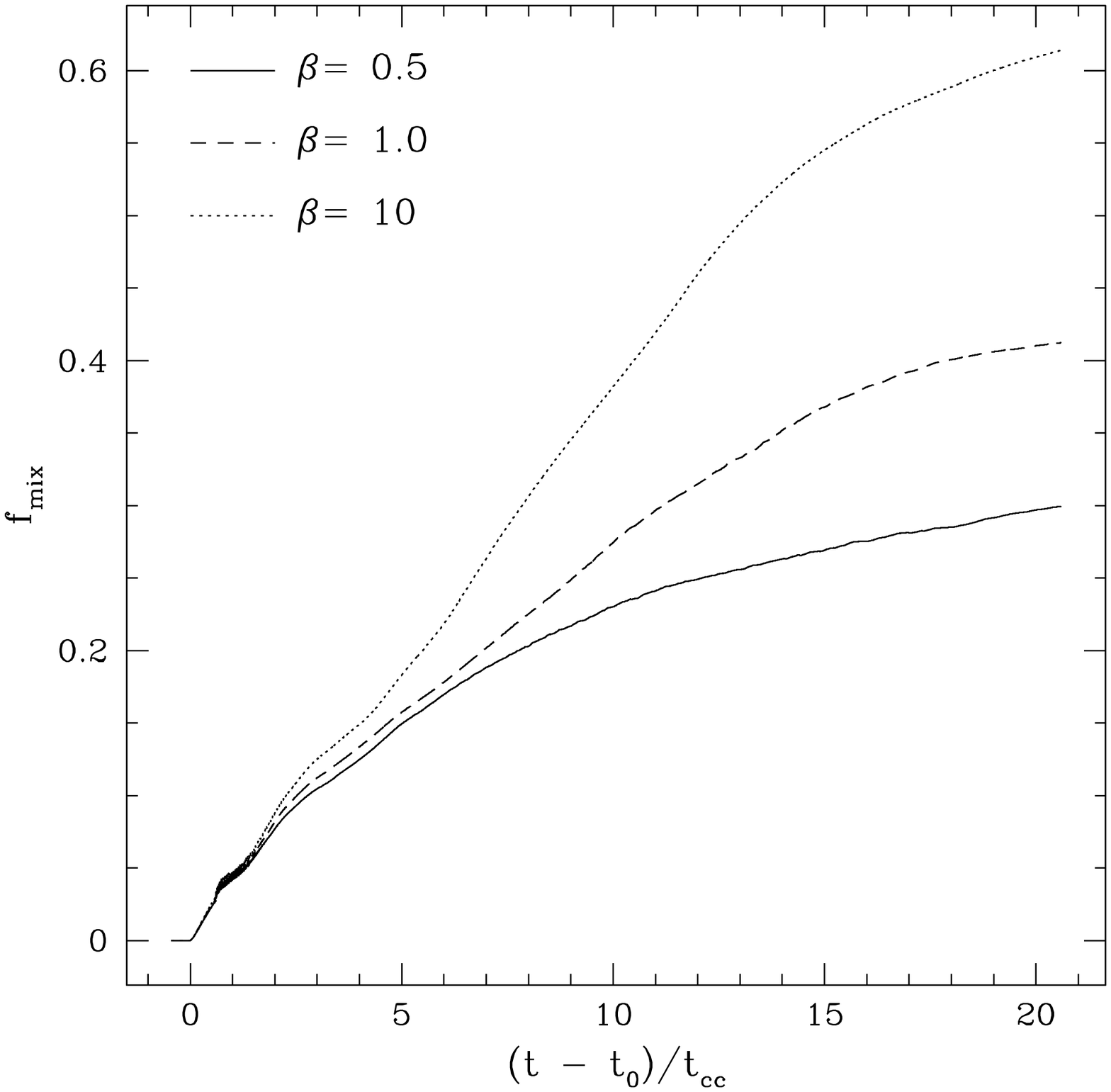}
\figcaption{Evolution of the mixing fraction for perpendicular shocks with
different initial magnetic field strength, measured by the initial $\beta$.
}
\end{figure}

\begin{figure}[t]
\epsscale{1.0}
\plotone{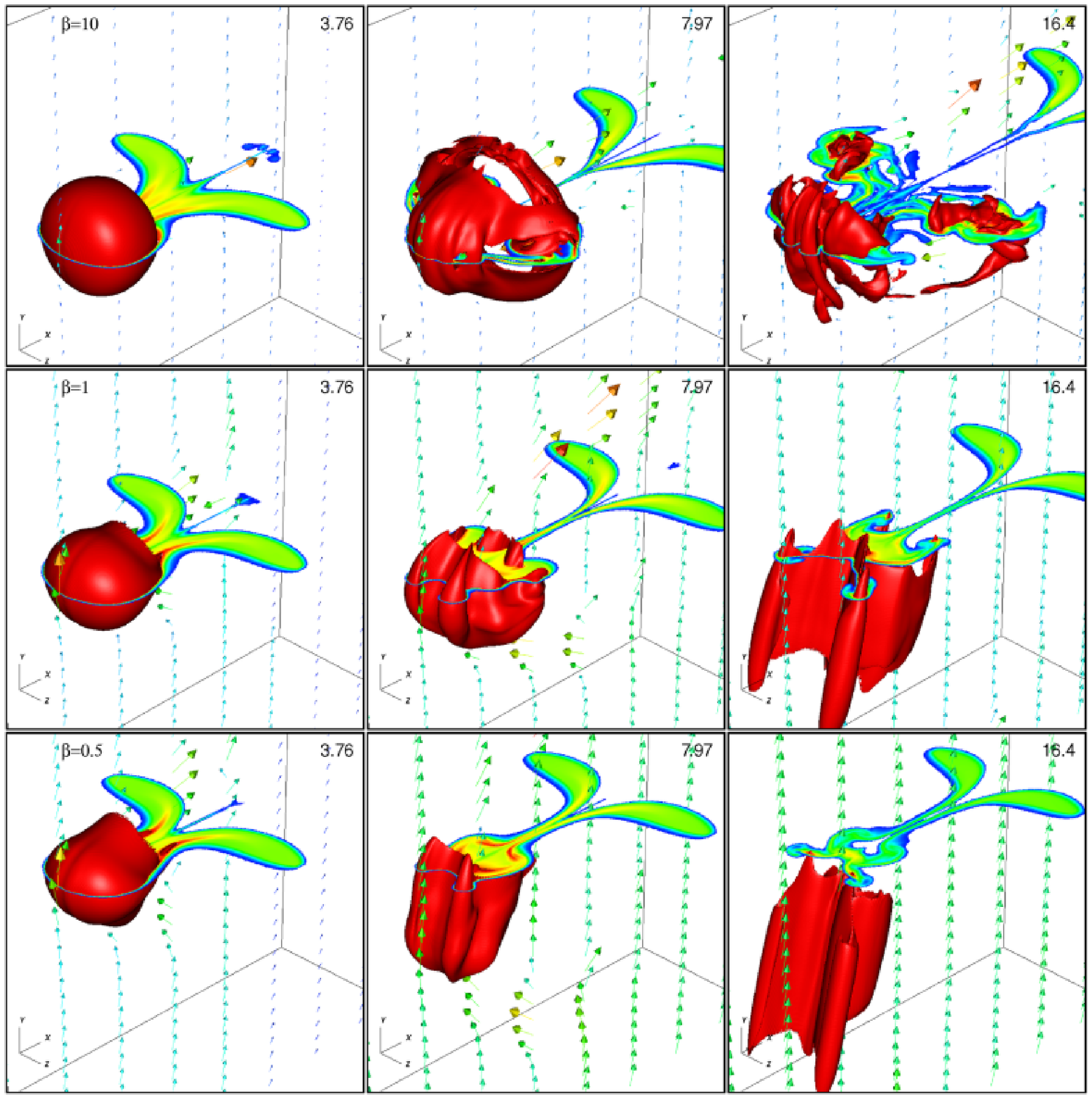}
\figcaption{Isosurface and horizontal slice at $y=0$ of the cloud mass density
$\rho C$, and magnetic field vectors (scaled and colored by 
$\parallel {\bf B} \parallel$)
on a vertical slice at $z=0$, for the
oblique shock simulations with an initial field that is
either weak ($\beta=10$, top row), equipartition ($\beta=1$, middle row),
or strong ($\beta=0.5$, bottom row).
}
\end{figure}

\begin{figure}[t]
\epsscale{0.4}
\plotone{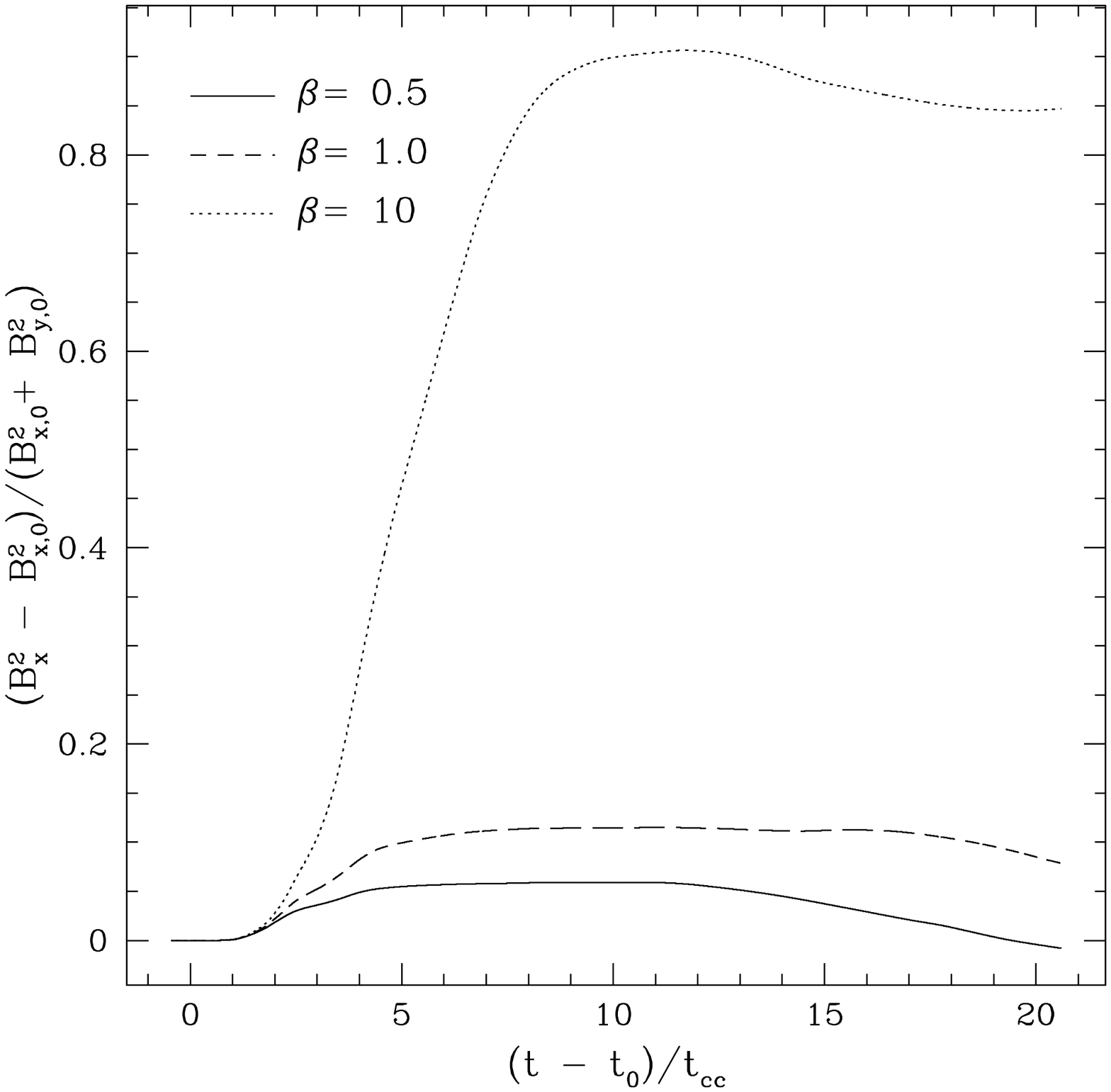}
\plotone{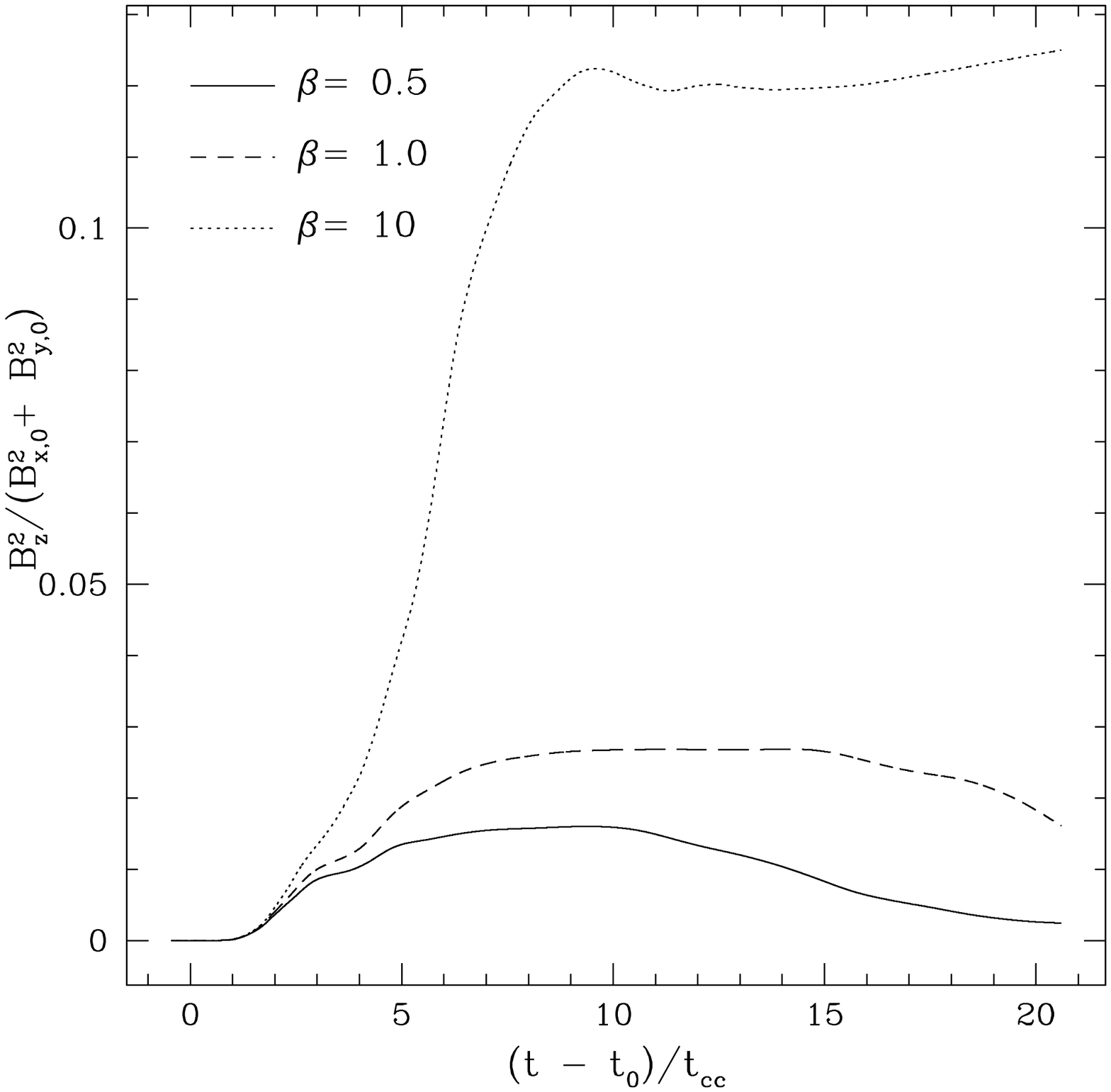}
\figcaption{Evolution of the change in magnetic energy density for different
$\beta$ for the oblique shock simulations, measured in units of the energy
density in the initial field $B_{x,0}^{2}+B_{y,0}^{2}$.
The evolution of $B_y^2$ is dominated by amplification due to
shock compression, and so is not shown.
}
\end{figure}

\begin{figure}[t]
\epsscale{1.0}
\plotone{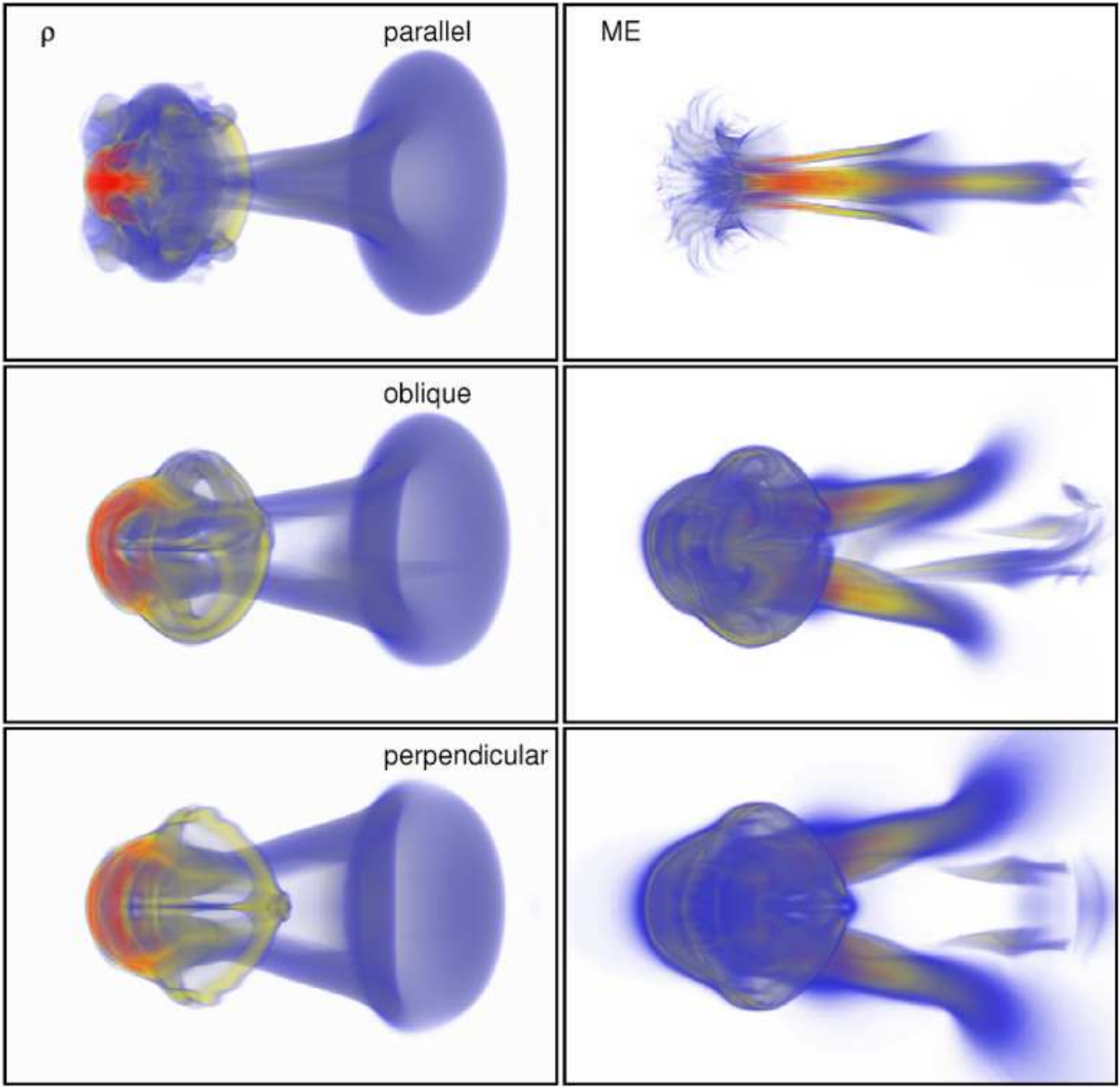}
\figcaption{Volumetric rendering of the cloud density $\rho C$ (left column)
and magnetic energy density $B^{2}$ (right column) for weak field ($\beta=10$)
parallel, perpendicular and oblique shocks.  Each plot is shown at
$(t-t_{0})/t_{cc} = 7.97$.
}
\end{figure}

\end{document}